\documentclass
[aps,showkeys,showpacs,pre,twocolumn,
superscriptaddress,amsmath,amssymb,amsfonts]{revtex4-1}
\usepackage{graphicx}
\usepackage{dcolumn}
\usepackage{bm}
\usepackage{amsmath}
\usepackage{mathrsfs}
\usepackage{hyperref}
\usepackage{xcolor}

\begin{document}

\title{Robustness analysis of bimodal networks in the whole range of degree correlation}

\author{Shogo Mizutaka}
\email[E-mail:]{mizutaka@ism.ac.jp}
\affiliation{School of Statistical Thinking, The Institute of Statistical Mathematics,
Tachikawa 190-8562, Japan}
\author{Toshihiro Tanizawa}
\email[E-mail:]{tanizawa@ee.kochi-ct.ac.jp}
\affiliation{Kochi National College of Technology,
200-1 Monobe-Otsu, Nankoku, Kochi 783-8508, Japan}

\date{\today}


\begin{abstract}
We present exact analysis of the physical properties of
bimodal networks specified by the two peak degree distribution
fully incorporating the degree-degree correlation between node connection.
The structure of the correlated bimodal network is uniquely
determined by the Pearson coefficient of the degree correlation, keeping its degree distribution fixed.
The percolation threshold and the giant component fraction of the correlated bimodal network
are analytically calculated in the whole range of the Pearson coefficient from $-1$ to $1$
against two major types of node removal, which are the random failure and the degree-based targeted attack.
The Pearson coefficient for next-nearest-neighbor pairs is also calculated,
which always takes a positive value even when the correlation between nearest-neighbor pairs is negative.
From the results,
it is confirmed that the percolation threshold is a monotonically decreasing function
of the Pearson coefficient for the degrees of nearest-neighbor pairs increasing from $-1$ and $1$
regardless of the types of node removal.
In contrast,
the node fraction of the giant component for bimodal networks with positive degree correlation
rapidly decreases in the early stage of random failure,
while that for bimodal networks with negative degree correlation remains relatively large
until the removed node fraction reaches the threshold.
In this sense, bimodal networks with negative degree correlation are more robust
against random failure than those with positive degree correlation.
\end{abstract}

				    

\maketitle



\section{Introduction}
\label{sec:introduction}

The functions of complex systems in the real world represented by complex networks largely
depend on the global connectivity of the nodes.
Since the networks are embedded in the external environment,
they are not immune to the possibilities of node and/or link failure or their intentional removal.
Structural robustness of networks has been, therefore, one of the main issues in network science \cite{Cohen:2000vq,Callaway:2000vd,Cohen:2001hf,Valente:2004jq,Paul:2004br,Tanizawa:2005fd,Tanizawa:2006dn,Paul:2006cd}.
Among many possibilities of structure alteration to malfunction of a network,
random failure that uniformly occurs on every node and selective node removal from those with large degrees (degree-based targeted attack)
are the two main categories, to which this article also pay attention.

In the early stage of the study on network robustness, most works were only based on their degree distribution.
The effect of degree-degree correlation between node connection has, however, already been considered preliminarily by Newman in 2002 \cite{Newman:2002jj}.
In this work, Newman has firstly pointed out that the tendency of connection between nodes with the same degree (positive or assortative correlation)
increases the structural robustness against random failure of nodes,
while the tendency of connection between nodes with different degree (negative or disassortative correlation)
decreases the robustness \cite{Newman:2002jj,Goltsev:2008bf}.

Recently Schneider \textit{et al.} through numerical simulations and Tanizawa \textit{et al.} with strict analytical arguments found that
the most robust network structure with a given degree distribution under both possibilities of
random failure of nodes and degree-based targeted attack is a set of hierarchically connected random regular graphs
with the ratios of node numbers of each graph in accordance with the degree distribution \cite{Schneider:2011ip,Herrmann:2011hd,Tanizawa:2012hh}.
In this structure, which is commonly termed as the ``onion structure'' at present,
almost all nodes are connected with nodes with the same degree
and the degree correlation between nearest neighbor nodes is positively maximal.

Compared to this situation regarding to the cases of positive degree correlation,
robustness analysis of networks with negative degree correlation still remains at an elementary level,
though there are some works in the literature in this regard \cite{Shiraki:2010is,Tanizawa:2013fy} and
several researches concerning the lower bound of degree-degree correlations for scale-free networks
\cite{Menche:2010as,Litvak:2013un,Yang:2016lo}.
This is because no network model is proposed that can tune degree-degree correlations easily and
can be analysed.
Thinking that many networks with important functions in real life,
such as the World Wide Web (WWW) or neural networks, have negative degree correlation as pointed out by Newman,
it is preferable if we have more thorough theoretical analyses for networks with negative degree correlation.

The bimodal network is a class of networks consisted of two species of nodes with distinctive degrees.
Since it is a minimal non-trivial network model,
the bimodal network has extensively been studied in many works for the analysis of network robustness \cite{Paul:2004br,Tanizawa:2005fd,Paul:2006cd,Shiraki:2010is}.
In spite of this fact, no work seems to have pointed out
that the bimodal network can cover the whole range of degree correlation from maximally negative to maximally positive,
as far as the authors' knowledge.
In this paper, we elucidate the structural properties of bimodal networks in terms of degree-degree correlation between
node connection under the condition of fixed degree distribution through analytical arguments.
There we will see that, though it is true that the threshold for remaining node fraction against
both random failure and degree-based targeted attack always decreases
as the degree correlation of the network increases to the positive direction,
the collapse of the giant component occurs much faster in the early stage of random failure.
In other words, the strong assortativity in bimodal networks makes networks more fragile against random failure.

The paper is organized as follows.
In Sec.~\ref{sec:theory}, we set the notations and calculate several important quantities
such as the degree distribution, the branching probability, the Pearson coefficient for
nearest neighbor nodes. The Pearson coefficient for next-nearest neighbor nodes is also calculated.
All calculations are performed exactly.
In Sec.~\ref{sec:Percolation}, the percolation threshold for random failure of nodes
and that for degree-based targeted attack are calculated.
In Sec.~\ref{sec:R_measure}, we discuss the network robustness in terms of the measure
representing the collapse of the giant component.
In Sec.~\ref{sec:discussions} we summarize the results and conclude the paper.

\section{Properties of correlated bimodal networks}
\label{sec:theory}
\begin{figure}[tttt]
\begin{center}
\includegraphics[width=0.35\textwidth]{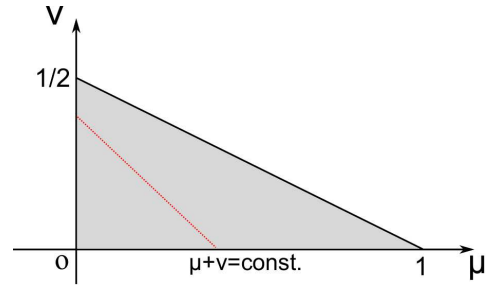}
\caption{Sketch of the allowable region in the $\mu$-$\nu$ plane (gray region). 
The red dotted line represents a set of bimodal networks with
the same degree distribution $P(k)$ and the different values of the joint probability \(Q(q_1, q_2)\).
}
\label{fig1}
\end{center}
\end{figure}

For the calculation, we introduce the joint 
probability \(Q(q_1, q_2)\) that a randomly chosen edge
has a node with \(q_1\) extra edges at one end
and a node with \(q_2\) extra edges at the other end.
Notice that the network is undirected
and that the chosen edge is not counted in this probability.
For a bimodal network of degree \(m\) and degree \(K\) with
\(m < K\), the probability \(Q(q_1, q_2)\) can be defined as
\begin{equation}
\label{eq:1}
\begin{pmatrix}
Q(\tilde{m}, \tilde{m}) & Q(\tilde{m}, \tilde{K}) \\
Q(\tilde{K}, \tilde{m}) & Q(\tilde{K}, \tilde{K})
\end{pmatrix}
=
\begin{pmatrix}
1 - 2\nu - \mu & \nu \\
\nu            & \mu
\end{pmatrix},
\end{equation}
where $\tilde{m} = m - 1$ and $\tilde{K} = K - 1$.
The two parameters,
$\mu (= Q(\tilde{K}, \tilde{K}))$ and $\nu (= Q(\tilde{K}, \tilde{m}))$, 
taking the values in the ranges, $0\le \mu \le 1$ and $0\le \nu \le 1/2$,
completely determine the network structure.
From Eq.~(\ref{eq:1}), the probability $Q(q)$ that a randomly chosen edge connects a node having
$q$ edges other than the chosen edge is given by
\begin{align}
\label{eq:3}
Q(\tilde{m}) &= Q(\tilde{m}, \tilde{K}) + Q(\tilde{m}, \tilde{m}) = 1 - (\mu + \nu), \\
Q(\tilde{K}) &= Q(\tilde{K}, \tilde{m}) + Q(\tilde{K}, \tilde{K}) = \mu + \nu.
\end{align}
Since $Q(\tilde{k})$ is related to the degree distribution, $P(k)$,
via the equations,
\begin{align}
\label{eq:4}
Q(\tilde{m}) &= \frac{m P(m)}{m P(m) + K P(K)}, \\
Q(\tilde{K}) &= \frac{K P(K)}{m P(m) + K P(K)}
\end{align}
for the bimodal network,
the degree distribution becomes
\begin{align}
\label{eq:5}
P(m) &= \frac{K \left\{ 1 - \left( \mu+\nu \right) \right\} }{m \left( \mu+\nu \right) + K \left\{1 - \left( \mu+\nu \right)  \right\} }, \\
P(K) &= \frac{m \left( \mu+\nu \right)}{m \left( \mu+\nu \right) + K \left\{1 - \left( \mu+\nu \right)  \right\} },
\end{align}
with the average degree
\begin{equation}
\label{eq:kav}
\langle k\rangle   = \frac{mK}{m(\mu + \nu)+K\left\{ 1-(\mu + \nu) \right\} }.
\end{equation}
Notice that the quantities, Eqs.~\eqref{eq:5}-\eqref{eq:kav}, are determined by the sum
$\mu + \nu (= Q(\tilde{K}))$, which we denote by $c$, indicating that, by fixing the sum,
we are able to fix the average degree, or equivalently, the total number of the edges.
The allowable area in the $\mu$-$\nu$ plane is shown in
Fig.~\ref{fig1}. 
The red dotted line in the figure indicates a set of parameter values, $\mu$ and $\nu$,
with a fixed sum $c = \mu + \nu$
representing the bimodal networks with the same degree distribution, $P(k)$,
and the different values of the joint probability, \(Q(q_1, q_2)\).

\subsection{Correlation between the nearest neighbor degrees}

The Pearson coefficient defined by
\begin{equation}
\label{eq:7}
r = \frac{ \langle q_1 q_2 \rangle - \langle q \rangle^2}{\langle q^2 \rangle - \langle q \rangle^{2}}
\end{equation}
is a commonly used measure for the degree-degree correlation between two extra edges,
$q_1$ and $q_2$, of both ends of a randomly selected edge.
Since
\begin{align}
\label{eq:10}
\langle q_1 q_2 \rangle &= \sum_{q_{1}, q_{2}}q_{1}q_{2}Q(q_{1},q_{2}) = \tilde{m}^2 \left( 1 - 2\nu - \mu \right) + 2\tilde{m} \tilde{K} \nu + \tilde{K}^2 \mu, \\
\langle q \rangle &= \sum_{q}qQ(q) = \tilde{m} \left(1 - c \right) + \tilde{K} c,
\end{align}
and
\begin{equation}
\langle q^2 \rangle = \sum_{q}q^{2}Q(q) = \tilde{m}^2 \left(1 - c \right)  + \tilde{K}^2 c,
\end{equation}
for bimodal networks,
we are able to obtain the analytical form of the Pearson coefficient, which becomes
\begin{equation}
\label{eq:14}
r = \frac{(1- \mu -2\nu)\mu - \nu^2}{\left(1 - c \right)c }
\end{equation}
with $c = \mu + \nu$.
It should be noted that we can choose any value of the correlation
coefficient $r$ without changing the degree distribution $P(k)$ or $\langle k \rangle$
by tuning $\mu$ and $\nu$ with keeping $c = \mu + \nu$ constant.
%
\begin{figure}[tttt]
\begin{center}
\includegraphics[width=0.48\textwidth]{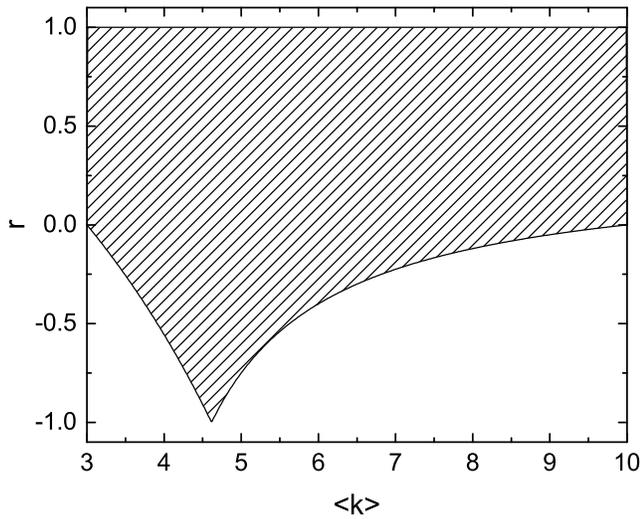}
\caption{Domain of the Pearson's coefficient $r$ as a function of
the average degree $\langle k\rangle$. The result is
for bimodal networks with $m=3$ and $K=10$.
}
\label{fig2_new}
\end{center}
\end{figure}

Figure~\ref{fig2_new} shows
the domain of the Pearson correlation coefficient $r$
in terms of the average degree $\langle k \rangle$
for bimodal networks with $K=10$ and $m=3$.
The range of the average degree is $3< \langle k\rangle <10$.
While all positive values within the range $(0, 1)$ can be realized 
for any value of the average degree, 
the domain of allowable negative values of $r$ is quite narrow.
The whole values of $r$ within the range $[-1, 1)$ can be realized only at a special value of $c = 1/2$, 
which corresponds to $\langle k \rangle = 60/13 = 4.615$. 

Strictly speaking, the network of the maximum Pearson coefficient (\(r = 1\))
consists of two separated random regular graphs with degree \(K\) and \(m\).
In this paper, we only consider the robustness of entirely connected networks.
The maximum value of the Pearson coefficient, \(r = 1\), is therefore to be realized as a limiting
value of \(r\) approaching one from below (\(r \to 1 - 0\)).
In this limit, the structure of the bimodal network consists of two random regular networks
connected by a few number of edges, 
which is known as the onion-like structure \cite{Schneider:2011ip,Herrmann:2011hd,Tanizawa:2012hh} 
and is always possible even if the degree inhomogeneity is weak.
On the other hand, in the construction of networks with negative values of the Pearson coefficient, 
the total number of edges from nodes of \(K\) degree should be equal to
the total number of edges from nodes of \(m\) degree: \(K P(K) = m P(m)\).
This condition poses a severe restriction in connecting nodes with different degrees
and leads to the narrow region of negative values in Fig.~\ref{fig2_new}.

\subsection{Correlation between the next-nearest neighbor degrees}
\label{sec:second_nearest}

Let \(Q_2(\tilde{k}_1, \tilde{k}_2)\) be the probability
that a randomly chosen node pair sharing a common neighbor in between
has $\tilde{k}_1$ and $\tilde{k}_2$ edges.
It should be noticed that two edges from the common neighbor
leading to both nodes of the pair are not included in
the edge counts, \(\tilde{k}_1\) and \(\tilde{k}_2\).
The probability is given by
\begin{widetext}
\begin{equation}
\label{eq:Q2_gen}
Q_2(\tilde{k}_1, \tilde{k}_2) = \frac{\sum_k k(k-1)P(k) Q(\tilde{k}_1|\tilde{k})
Q(\tilde{k}_2|\tilde{k})\left\{1-Q(\tilde{k}_1,\tilde{k}_2)\right\}}
{\sum_k\sum_{\tilde{k}_1, \tilde{k}_2} k(k-1)P(k) Q(\tilde{k}_1|\tilde{k})
Q(\tilde{k}_2|\tilde{k})\left\{1-Q(\tilde{k}_1,\tilde{k}_2)\right\}},
\end{equation}
\end{widetext}
where \(Q(\tilde{k}'|\tilde{k})\) is the branching probability that an edge
emanating from a node of degree \(k\) reaches a node of degree \(k'\) 
defined by $Q(\tilde{k}'|\tilde{k}) = {Q(\tilde{k}, \tilde{k}')}/
{Q(\tilde{k})}$.
If the network structure is locally tree-like as assumed in this paper,
the total number of node pairs of sharing a common neighbor is proportional to
\(
\sum_{k}k(k-1)P(k)=\langle k^2 \rangle-\langle k\rangle
\)
and hence the probability \(Q_2(\tilde{k}_1, \tilde{k}_2)\) is given by
\begin{equation}
\label{eq:Q2}
Q_2(\tilde{k}_1, \tilde{k}_2) = \frac{\sum_k k(k-1)P(k) Q(\tilde{k}_1|\tilde{k}) Q(\tilde{k}_2|\tilde{k})}{\langle k^2 \rangle-\langle k\rangle}.
\end{equation}
In the present bimodal case, the non-zero branching probabilities are
\begin{align}
\label{eq:19}
Q( \tilde{m} | \tilde{m} ) &= \frac{1-2\nu-\mu}{1-c}, \\
Q( \tilde{K} | \tilde{m} ) &= \frac{\nu}{1- c }, \\
Q( \tilde{m} | \tilde{K} ) &= \frac{\nu}{c},
\end{align}
and
\begin{equation}
Q( \tilde{K} | \tilde{K} ) = \frac{\mu}{c}.
\end{equation}
With the probability $Q_2(\tilde{k}_1, \tilde{k}_2)$, we can
calculate the Pearson correlation coefficient $r_{2}$ for the degrees of
next-nearest neighbors:
\begin{equation}
\label{eq:20}
r_2 = \frac{{\langle q_1 q_2 \rangle}_{2} - {{\langle q \rangle}_2}^2}{{\langle q^2 \rangle}_{2} - {{\langle q \rangle}_2}^2},
\end{equation}
where
\begin{align}
\label{eq:21}
{\langle q_1 q_2 \rangle}_2 &= \sum_{q_1, q_2} q_1 q_2 Q_2(q_1, q_2), \\
{\langle q \rangle}_2       &= \sum_{q, q_2} q~Q_2(q, q_2),
\end{align}
and
\begin{equation}
{\langle q^2 \rangle}_2      = \sum_{q, q_2} q^2 Q_2(q, q_2).
\end{equation}
\begin{figure}[tttt]
\begin{center}
\includegraphics[width=0.48\textwidth]{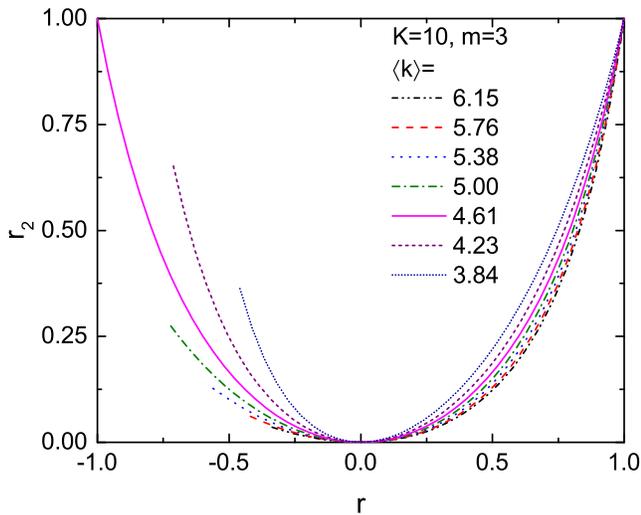}
\caption{Plot for the values of the Pearson correlation coefficient $r_2$ for the degrees of
next-nearest neighbors as a function of the Pearson coefficient $r$ for
the degrees of nearest neighbors. All lines are the results for bimodal
networks with degree $m=3$ and degree $K=10$ and the lines
from bottom to top correspond to the results for the average
degree $\langle k\rangle\approx 6.15$, $5.76$, $5.38$, $5.00$, $4.61$,
$4.23$, and $3.84$, respectively.
}
\label{fig3}
\end{center}
\end{figure}
Figure~\ref{fig3} shows the Pearson correlation coefficient $r_2$
for the next-nearest neighbors as a function of the Pearson coefficient $r$ for the nearest neighbors.
It is noteworthy that
$r_2$ of the correlated bimodal networks 
never takes negative values irrespective to the values of $r$.
For the cases of $r > 0$,
it is reasonable, since
the entire graph contains subgraphs consisting of only nodes of $m$-degree
or $K$-degree that give positive contribution to $r_2$.
Even for the cases of $r < 0$, however,
the next-nearest neighbors have tendency to be of the same degree,
since there are only two degrees, $m$ and $K$, in bimodal networks.
The value of $r_2$ for $r = -1$ reaches its maximum $r_2 = 1$ only when $c = \mu + \nu = 1/2$.
The physical picture for $r_2 = 1$ for $r = -1$ is, however, different from
that of the case $r_2 = 1$ for $r = 1$.
At $r = 1$, the network is completely separated into two random regular graphs of
degree-$m$ and degree-$K$,
while at $r = -1$ the entire network forms a single connected component
with nodes of degree $m$ and nodes of  degree $K$ adjacent with each other,
which implies that the next-nearest neighbors have the same degree.

\section{Percolation threshold}
\label{sec:Percolation}

For the networks with tree-like structure,
the calculation method for various physical quantities,
such as the percolation threshold and the node fraction of the largest
connected component, has been well-established
even in the cases with arbitrary degree-degree correlation and the ways
of node removal \cite{Goltsev:2008bf,Tanizawa:2012hh}.
Let $x_k$ be the probability that
a randomly chosen edge from a $k$-degree node does not lead to the giant component.
These $x_k$ are the solutions of the simultaneous equations
\begin{equation}
\label{eq:xk}
x_k  = 1- \sum_{k'}b_{k'}Q(\tilde{k}'|\tilde{k}) + \sum_{k'}b_{k'}Q(\tilde{k}'|\tilde{k})(x_{k'})^{\tilde{k}'},
\end{equation}
where $b_{k}$ is the remaining fraction of $k$-degree nodes
and $p$ is the total remaining fraction of nodes, i.e.,
$p=\sum_{k}b_{k}P(k)$.
With these $x_k$,
the node fraction of the giant component, $S$, is given by
\begin{equation}
\label{eq:Sp}
S(p)   = p - \sum_{k}b_{k}P(k)(x_{k})^{k}.
\end{equation}

Below the percolation threshold ($p < p_{\rm c}$),
all solutions of Eq.~(\ref{eq:xk}) {are trivial, $x_{k}=1$,
for any $k$, which means that the network does not contain
the giant component ($S=0$). Since the giant component emerges
in the network at criticality, at least one of
$x_{k}$'s takes a value slightly smaller than unity at the point.
Rewriting as
$y_{k}=1-x_{k}$, we expand Eq.~(\ref{eq:xk}) near the
critical point in terms of $y_k$'s
and obtain the equation
\begin{equation}
\label{eq:eigeneqyk}
y_k = B_{kk'} y_{k'}
\end{equation}
taking into account only the linear
term of the expanded equation, where
\begin{equation}
B_{kk'}=b_{k'}\tilde{k'}Q(\tilde{k'}|\tilde{k})
\end{equation}
is the branching matrix.
Equation \eqref{eq:eigeneqyk} implies that
the largest eigenvalue of the branching matrix $B_{kk'}$
become unity at criticality \cite{Goltsev:2008bf,Tanizawa:2012hh}.
Therefore the critical remaining node fraction $p_{\text{c}}$ for the 
bimodal network is determined by
\begin{gather}
\label{eq:b_mt}
\begin{vmatrix}
b_m \tilde{m} Q(\tilde{m}|\tilde{m}) - 1 & b_K \tilde{K} Q(\tilde{K}|\tilde{m}) \\
b_m \tilde{m} Q(\tilde{m}|\tilde{K})     & b_K \tilde{K} Q(\tilde{K}|\tilde{K}) - 1
\end{vmatrix}
= 0, \\
p_{\text{c}} = b_m P(m) + b_K P(K)
\end{gather}
where the left-hand side of Eq.~(\ref{eq:b_mt}) represents the determinant.

\subsection{Percolation threshold for random failure}
\label{sec:orgheadline6}
In the case of the random failure, all remaining fractions
$b_{k}$'s for $k$-degree nodes take the same value.
In this case, Eq.(\ref{eq:b_mt}) becomes
\begin{equation}
\label{eq:22}
\begin{vmatrix}
p_{\rm c} \tilde{m} Q(\tilde{m}|\tilde{m}) - 1 & p_{\rm c} \tilde{K} Q(\tilde{K}|\tilde{m}) \\
p_{\rm c} \tilde{m} Q(\tilde{m}|\tilde{K})     & p_{\rm c} \tilde{K} Q(\tilde{K}|\tilde{K}) - 1
\end{vmatrix}
= 0,
\end{equation}
where $p_{\rm c}$ is the percolation threshold for the random
node removal. Therefore, the threshold $p_{\rm c}$ is the root of the
equation
\begin{equation}
\label{eq:24}
a {p_{\rm c}}^2 - b p_{\rm c} + 1 = 0
\end{equation}
with
\begin{align}
\label{eq:26}
a &= \tilde{m} \tilde{K} \left( Q(\tilde{m}|\tilde{m}) Q(\tilde{K}|\tilde{K}) - Q(\tilde{K}|\tilde{m}) Q(\tilde{m}|\tilde{K}) \right) = r \: \tilde{m} \tilde{K}, \\
b &= \tilde{m} Q(\tilde{m}|\tilde{m}) + \tilde{K} Q(\tilde{K}|\tilde{K}),
\end{align}
where $r$ is the Pearson coefficient for the degree correlation between nearest-neighbor node pairs.
Thus, for \(r \neq 0\)
\begin{equation}
\label{eq:11}
p_{\rm c} = \frac{b - \sqrt{b^2 - 4 a}}{2 a}
\end{equation}
and for \(r = 0\)
\begin{equation}
\label{eq:28}
p_{\rm c} = \frac{1}{b} = \frac{1}{ \tilde{m} Q(\tilde{m}|\tilde{m}) + \tilde{K} Q(\tilde{K}|\tilde{K}) }.
\end{equation}
Notice that $\lim_{r \to 0} p_{\text{c}} = 1/b$ as so it should be.
Figure~\ref{fig4} shows the percolation threshold $p_{\rm c}$ for
random failure as a function of the Pearson coefficient $r$. The percolation
threshold $p_{\rm c}$ are decreasing functions of $r$
regardless to the average degree.
The result shows that the positive degree correlation
always gives smaller values of $p_{\text{c}}$ against random failure for bimodal networks.
In addition, the decreasing rate of $p_{\rm c}$ in terms of $r$ becomes large as the average degree decreases.
%
\begin{figure}[tttt]
\begin{center}
\includegraphics[width=0.48\textwidth]{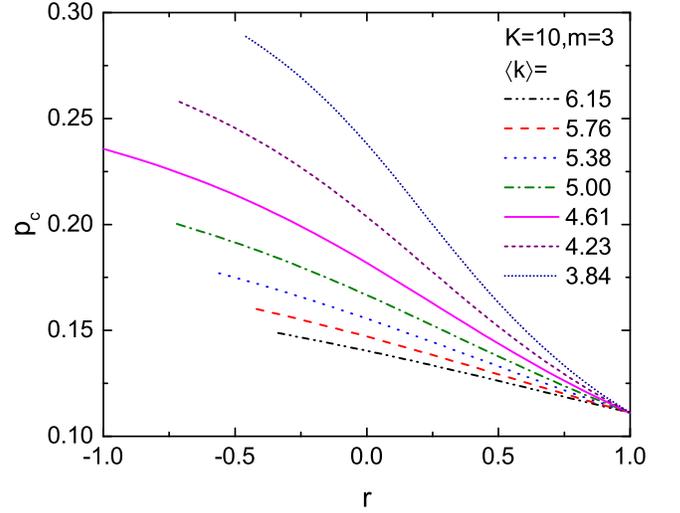}
\caption{The percolation threshold $p_{\rm c}$ for random failure as a function of the Pearson coefficient $r$
for various values of the average degree $\langle k \rangle$.
}
\label{fig4}
\end{center}
\end{figure}

\subsection{Percolation threshold for degree based targeted attack}
Next, we evaluate the percolation threshold for degree based targeted attack.
In this attack strategy for bimodal networks,
the percolation transition takes place when either
(i) only a part of $K$-degree nodes are removed ($b_m = 1$ and $0 \le b_K < 1$) or
(ii) all $K$-degree nodes and a part of $m$-degree nodes are removed ($0 \le b_m < 1$ and $b_K = 0$).
By substituting $b_{m}=1$ and $b_{K}=0$ into Eq.~(\ref{eq:b_mt}),
we obtain the condition
\begin{equation}
\label{eq:29}
\nu = \left(1 - c \right) \left( 1- \frac{1}{\tilde{m}} \right)
\end{equation}
with the percolation threshold $p_{\text{c}} = P(m)$.
For $\nu > \left(1 - c \right) \left(1- \tilde{m}^{-1} \right)$ (Case (i)),
then the percolation threshold $p_{\rm c}$ is given by
\begin{equation}
\label{eq:30}
p_{\rm c}= P(m)+b_{K}P(K),
\end{equation}
where
\begin{equation}
\label{eq:bK}
b_{K}=\frac{\tilde{m} Q(\tilde{m}|\tilde{m}) -1 }{ \tilde{m}\tilde{K}r-\tilde{K}Q(\tilde{K}|\tilde{K})}.
\end{equation}
On the other hand, we obtain the percolation threshold
\begin{equation}
\label{eq:300}
p_{\rm c} = \frac{P(m)}{\tilde{m} Q(\tilde{m}|\tilde{m}) }
\end{equation}
for $\nu < \left(1 - c \right) \left(1- \tilde{m}^{-1} \right)$ (Case (ii)).
The dependence of the percolation threshold
$p_{\rm c}$ on the Pearson coefficient $r$ is shown in Fig.~\ref{fig5}.
The threshold $p_{\rm c}$ for targeted attack
is also a decreasing function of $r$ as the threshold $p_{\rm c}$ for random failure.
These tendencies have been reported by several works \cite{Newman:2002jj,Newman:2003fh,Vazquez:2003fq,Goltsev:2008bf}.
The behavior of $p_{\rm c}$ is separated at the value of $\nu$ represented in Eq.~\eqref{eq:29}.
While for $\nu>\left(1 - c \right) \left(1- \tilde{m}^{-1} \right)$
which corresponds to Case (i), $b_m = 1$ and $0 \le b_K < 1$,
the slope of $p_{\rm c}$ in terms of $r$ is gentle,
while for $\nu<\left(1 - c \right) \left(1- \tilde{m}^{-1} \right)$,
which corresponds to Case (ii), $0 \le b_m < 1$ and $b_K = 0$,
the value of $p_{\rm c}$ is sensitive to the change of the degree correlation $r$.
This trend do not depend on values of $m$ and $K$ and the slope for
$\nu>\left(1 - c \right) \left(1- \tilde{m}^{-1} \right)$ becomes
more gentle with increasing the value of $K$ as seen from
Eq.~(\ref{eq:bK}).

\begin{figure}[tttt]
\includegraphics[width=.48\textwidth]{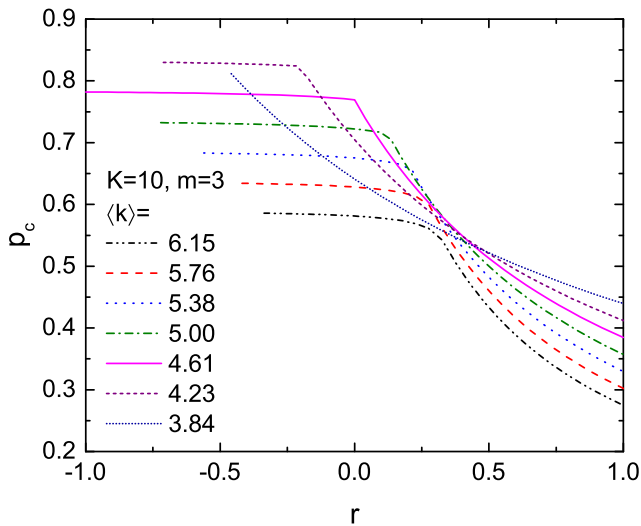}
\caption{Pearson's $r$ dependence of the percolation
threshold $p_{\rm c}$ for targeted attack. Legend is the same as Fig.~\ref{fig3}.
}
\label{fig5}
\end{figure}

Either for random failure or for targeted attack,
the value of the percolation threshold decreases as the Pearson coefficient increases from $-1$ to $1$.

\section{Network robustness in terms of the giant component collapse}
\label{sec:R_measure}
In Sec.~\ref{sec:Percolation}, we discussed the robustness of
correlated bimodal networks in terms of the percolation thresholds.
The percolation threshold is, however, not the only index for the network robustness.
Even if the network has a small value of threshold,
the giant component might rapidly collapse in the early stage of node removal.
To take this possibility into account, Schneider et al. recently introduced
a new measure $R$ defined by
\begin{equation}
R=\int_{0}^{1}S(p)dp,
\end{equation}
where $S(p)$ is the node ratio of the giant component to the initial network
when the remaining node fraction is $p$ \cite{Schneider:2011ip,Herrmann:2011hd}.
Since $S(p) \le p$,
the values of $R$ reside in the range $0 \le R \le 1/2$.
The new measure
$R$ contains information of the relative size of the giant component for any $p$
and reflects the sensitivity of the change of the giant component size for the node removal.
Figure~\ref{fig6} shows the relative size $S(p)$ of the
giant component as a function of the remaining node fraction $p$.
Figure~\ref{fig6} (a) is the plot for random node removal
and Fig.~\ref{fig6} (b) for targeted attack.
The solid and broken lines in the figure are the values obtained
by solving Eqs.~(\ref{eq:xk}) and (\ref{eq:Sp}) numerically
and the symbols are from numerical simulations.
The agreement of the values from numerical simulations with theoretical values is
almost perfect.
As seen in Fig.~\ref{fig6}~(a), 
$S(p)$ for the networks with strong positive correlation
(red solid line and red open circle) against random failure
decreases rapidly as the decrease of $p$, which implies that
$R$ for this kind of network takes a small value
even if the percolation threshold for this case is the smallest,
while $S(p)$ for the networks with negative degree correlation decreases more slowly.
In other words, the bimodal networks with strong positive degree correlation
are fragile against random failure in terms of $R$,
which seems contrary to the common knowledge that the positive degree correlation makes networks robust
at first thought.
It should be noted that the rapid drop in the curve of \(S(p)\) for the case of positive degree correlation shown
in Fig.~\ref{fig6}(a) can be interpreted as an indication of double percolation transition similar to the one reported in \cite{Colomer:2014},
since the network structure for the strong positive degree correlation consists of two random regular graphs of degree \(m\) and \(K\) connected
by a few number of edges.
Actually, we have confirmed numerically that the mean cluster size in this case has two peaks
at the points of two percolation thresholds corresponding to the collapses of the components of $m$-(or \(K\)-)degree nodes.
For targeted attack, on the other hand, 
the bimodal networks with strong positive degree correlation 
have the largest value of $R$, which has been pointed out numerically and analytically \cite{Schneider:2011ip,Herrmann:2011hd,Tanizawa:2012hh}.
\begin{figure}[tttt]
\begin{center}
\includegraphics[width=0.48\textwidth]{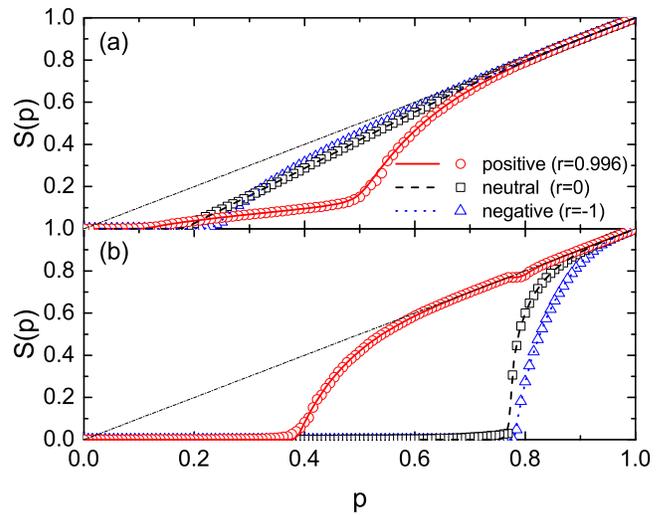}
\caption{Plots of the relative size $S(p)$ of the giant component as a function of the remaining node fraction $p$
(a) for random failure and (b) for targeted attack.
The lines are the results obtained by solving Eqs.~(\ref{eq:xk})
and (\ref{eq:Sp}) numerically. Symbols are the results from numerical simulations
under the condition of the number of nodes $N=13000$
for bimodal networks with $m=3$, $K=10$ and $c=0.5$ (i.e., $\langle k\rangle \approx 4.61$).
The broken lines, $S(p)=p$, representing the maximum robustness are also plotted for the guide to the eye
in the both plots.
}
\label{fig6}
\end{center}
\end{figure}

In Fig.~\ref{fig7}, we display the robustness measure $R$ as a function
of the Pearson coefficient $r$. The robustness measure $R$ is a slowly decreasing
function of $r$ for random failure (black filled square), which
implies that the negative degree correlation makes bimodal networks
robust against random failure and keeps the giant component size as large as possible.
For targeted attack, $R$ is an increasing function of
$r$ (red filled circle). In this case, the bimodal networks with negative degree correlation are
fragile against targeted attack in both contexts of the robustness measure
$R$ and the percolation threshold $p_{\rm c}$,
while the positive degree correlation makes bimodal networks robust against the targeted attack.
The physical reason that
the robustness $R$ for random failure is lower than that for targeted attack in the region of large $r$ (\(\gtrsim 0.9\)) is the following.
In bimodal networks with large \(r (\gtrsim 0.9)\), the giant component is almost separated into two random regular graphs of degree \(m\) and \(K\)
with a larger number of \(m\)-degree nodes because of the condition for matching edges: \( K P(K) = m P(m)\).
For random failure, the component of degree \(m\) collapses earlier inducing the rapid shrink in the size of the giant component.
For targeted attack, on the other hand, $K$-degree nodes are removed firstly
and the component of \(m\)-degree nodes is intact, which retains the size of the giant component.
The inset of Fig.~\ref{fig7} shows the $r$ dependence of the sum $R_{\rm tot}$ of the
robustness measure for random failure and that for targeted attack.
The optimal value of the Pearson coefficient for both random failure and targeted attack in
terms of $R_{\rm tot}$ is in the range of the positive values.

The tendencies of the robustness described above does not depend on the parameters $m$ and $K$.
If we choose larger $K$,
the node fraction of the giant component $S(p)$ for random failure
decreases much more rapidly according to the decrease of the Pearson coefficient $r$,
which makes the value of $R$ much smaller.

\begin{figure}[tttt]
\begin{center}
\includegraphics[width=0.48\textwidth]{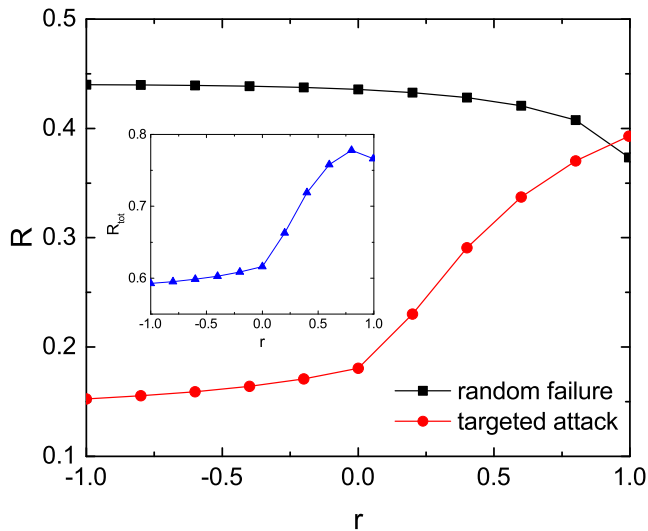}
\caption{Plot of the robustness measure $R$ as a function of
the Pearson coefficient $r$ for random failure (black filled square) and
targeted attack (red filled circles). The inset displays the $r$ dependence of the
total sum $R_{\rm tot}$ of the robustness measure for random failure
and that for targeted attack.
}
\label{fig7}
\end{center}
\end{figure}

\section{Summary}
\label{sec:discussions}

In this paper,
we investigate the structural robustness of correlated bimodal networks
under the possibilities of random failure of nodes and degree-based targeted attack on nodes.
Several important structural properties are calculated exactly as a function of
the Pearson coefficient for degree correlation between nearest-neighbor nodes
in the whole range of the coefficient keeping the degree distribution fixed.
The Pearson coefficient for degree correlation between next-nearest-neighbor nodes are also calculated,
which always takes positive values, regardless of the nearest neighbor correlation.
Next, we investigate the network robustness against random failure and
degree-based targeted attack 
with respect to the percolation threshold
$p_{\rm c}$ and the value $R$, which is the area under
the curve of the giant component fraction $S(p)$, as the robustness measure. 
Regardless of the type of node removal, the percolation threshold $p_{\rm c}$
for the networks with negative degree correlation always takes larger values than that of
networks with positive degree correlation,
which is consistent with the conclusion in existing works
\cite{Newman:2002jj,Newman:2003fh,Vazquez:2003fq,Goltsev:2008bf}.
In terms of the robustness measure $R$, however,
the giant component in the networks with positive degree correlation
collapses more rapidly in the early stage of random failure than
that in the networks with negative degree correlation.
In this sense, the assortative bimodal network is \emph{fragile}
against random failure of nodes,
which seems to be counter-intuitive at first glance.
This fragility comes from the fact that,
in assortative bimodal networks, the larger subgraph in the giant component
that consists of smaller degree nodes are not able to receive sufficient
support from larger degree nodes because of the tendency of making connectivity
between nodes of the same degree.
In this sense, for random failure of nodes, negative degree correlation
is favorable for the resilience of the connectivity in the giant component.
This effect of the negative degree correlation
on the giant component size seems to be persistent,
independent of the network structure as reported in Refs.~\cite{Newman:2002jj,Noh:2007cl}.
In contrast, the assortativity enhances the resilience of
the connectivity of the giant component against degree-based targeted attack
as investigated by Schneider et al.\ and Tanizawa et al.\ \cite{Schneider:2011ip,Herrmann:2011hd,Tanizawa:2012hh}.




\begin{thebibliography}{18}%
\makeatletter
\providecommand \@ifxundefined [1]{%
 \@ifx{#1\undefined}
}%
\providecommand \@ifnum [1]{%
 \ifnum #1\expandafter \@firstoftwo
 \else \expandafter \@secondoftwo
 \fi
}%
\providecommand \@ifx [1]{%
 \ifx #1\expandafter \@firstoftwo
 \else \expandafter \@secondoftwo
 \fi
}%
\providecommand \natexlab [1]{#1}%
\providecommand \enquote  [1]{``#1''}%
\providecommand \bibnamefont  [1]{#1}%
\providecommand \bibfnamefont [1]{#1}%
\providecommand \citenamefont [1]{#1}%
\providecommand \href@noop [0]{\@secondoftwo}%
\providecommand \href [0]{\begingroup \@sanitize@url \@href}%
\providecommand \@href[1]{\@@startlink{#1}\@@href}%
\providecommand \@@href[1]{\endgroup#1\@@endlink}%
\providecommand \@sanitize@url [0]{\catcode `\\12\catcode `\$12\catcode
  `\&12\catcode `\#12\catcode `\^12\catcode `\_12\catcode `\%12\relax}%
\providecommand \@@startlink[1]{}%
\providecommand \@@endlink[0]{}%
\providecommand \url  [0]{\begingroup\@sanitize@url \@url }%
\providecommand \@url [1]{\endgroup\@href {#1}{\urlprefix }}%
\providecommand \urlprefix  [0]{URL }%
\providecommand \Eprint [0]{\href }%
\@ifxundefined \urlstyle {%
  \providecommand \doi  [0]{\begingroup \@sanitize@url \@doi}%
  \providecommand \@doi [1]{\endgroup \@@startlink {\doibase
  #1}doi:\discretionary {}{}{}#1\@@endlink }%
}{%
  \providecommand \doi  [0]{doi:\discretionary{}{}{}\begingroup
  \urlstyle{rm}\Url }%
}%
\providecommand \doibase [0]{http://dx.doi.org/}%
\providecommand \Doi [0]{\begingroup \@sanitize@url \@Doi }%
\providecommand \@Doi  [1]{\endgroup\@@startlink{\doibase#1}\@@Doi}%
\providecommand \@@Doi [1]{#1\@@endlink}%
\providecommand \selectlanguage [0]{\@gobble}%
\providecommand \bibinfo  [0]{\@secondoftwo}%
\providecommand \bibfield  [0]{\@secondoftwo}%
\providecommand \translation [1]{[#1]}%
\providecommand \BibitemOpen [0]{}%
\providecommand \bibitemStop [0]{}%
\providecommand \bibitemNoStop [0]{.\EOS\space}%
\providecommand \EOS [0]{\spacefactor3000\relax}%
\providecommand \BibitemShut  [1]{\csname bibitem#1\endcsname}%
\bibitem [{\citenamefont {Cohen}\ \emph {et~al.}(2000)\citenamefont {Cohen},
  \citenamefont {Erez}, \citenamefont {ben-Avraham},\ and\ \citenamefont
  {Havlin}}]{Cohen:2000vq}%
  \BibitemOpen
  \bibfield  {author} {\bibinfo {author} {\bibfnamefont {R.}~\bibnamefont
  {Cohen}}, \bibinfo {author} {\bibfnamefont {K.}~\bibnamefont {Erez}},
  \bibinfo {author} {\bibfnamefont {D.}~\bibnamefont {ben-Avraham}}, \ and\
  \bibinfo {author} {\bibfnamefont {S.}~\bibnamefont {Havlin}},\ }\href@noop {}
  {\bibfield  {journal} {\bibinfo  {journal} {Phys. Rev. Lett.},\ }\textbf
  {\bibinfo {volume} {85}},\ \bibinfo {pages} {4626} (\bibinfo {year}
  {2000})}\BibitemShut {NoStop}%
\bibitem [{\citenamefont {Callaway}\ \emph {et~al.}(2000)\citenamefont
  {Callaway}, \citenamefont {Newman}, \citenamefont {Strogatz},\ and\
  \citenamefont {Watts}}]{Callaway:2000vd}%
  \BibitemOpen
  \bibfield  {author} {\bibinfo {author} {\bibfnamefont {D.~S.}\ \bibnamefont
  {Callaway}}, \bibinfo {author} {\bibfnamefont {M.~E.~J.}\ \bibnamefont
  {Newman}}, \bibinfo {author} {\bibfnamefont {S.~H.}\ \bibnamefont
  {Strogatz}}, \ and\ \bibinfo {author} {\bibfnamefont {D.~J.}\ \bibnamefont
  {Watts}},\ }\href@noop {} {\bibfield  {journal} {\bibinfo  {journal} {Phys.
  Rev. Lett.},\ }\textbf {\bibinfo {volume} {85}},\ \bibinfo {pages} {5468}
  (\bibinfo {year} {2000})}\BibitemShut {NoStop}%
\bibitem [{\citenamefont {Cohen}\ \emph {et~al.}(2001)\citenamefont {Cohen},
  \citenamefont {Erez}, \citenamefont {ben-Avraham},\ and\ \citenamefont
  {Havlin}}]{Cohen:2001hf}%
  \BibitemOpen
  \bibfield  {author} {\bibinfo {author} {\bibfnamefont {R.}~\bibnamefont
  {Cohen}}, \bibinfo {author} {\bibfnamefont {K.}~\bibnamefont {Erez}},
  \bibinfo {author} {\bibfnamefont {D.}~\bibnamefont {ben-Avraham}}, \ and\
  \bibinfo {author} {\bibfnamefont {S.}~\bibnamefont {Havlin}},\ }\href@noop {}
  {\bibfield  {journal} {\bibinfo  {journal} {Phys. Rev. Lett.},\ }\textbf
  {\bibinfo {volume} {86}},\ \bibinfo {pages} {3682} (\bibinfo {year}
  {2001})}\BibitemShut {NoStop}%
\bibitem [{\citenamefont {Valente}\ \emph {et~al.}(2004)\citenamefont
  {Valente}, \citenamefont {Sarkar},\ and\ \citenamefont
  {Stone}}]{Valente:2004jq}%
  \BibitemOpen
  \bibfield  {author} {\bibinfo {author} {\bibfnamefont {A.~X.~C.~N.}~\bibnamefont
  {Valente}}, \bibinfo {author} {\bibfnamefont {A.}~\bibnamefont {Sarkar}}, \
  and\ \bibinfo {author} {\bibfnamefont {H.~A.}~\bibnamefont {Stone}},\
  }\href@noop {} {\bibfield  {journal} {\bibinfo  {journal} {Phys. Rev.
  Lett.},\ }\textbf {\bibinfo {volume} {92}},\ \bibinfo {pages} {118702}
  (\bibinfo {year} {2004})}\BibitemShut {NoStop}%
\bibitem [{\citenamefont {Paul}\ \emph {et~al.}(2004)\citenamefont {Paul},
  \citenamefont {Tanizawa}, \citenamefont {Havlin},\ and\ \citenamefont
  {Stanley}}]{Paul:2004br}%
  \BibitemOpen
  \bibfield  {author} {\bibinfo {author} {\bibfnamefont {G.}~\bibnamefont
  {Paul}}, \bibinfo {author} {\bibfnamefont {T.}~\bibnamefont {Tanizawa}},
  \bibinfo {author} {\bibfnamefont {S.}~\bibnamefont {Havlin}}, \ and\ \bibinfo
  {author} {\bibfnamefont {H.~E.}\ \bibnamefont {Stanley}},\ }\href@noop {}
  {\bibfield  {journal} {\bibinfo  {journal} {Eur. Phys. J. B},\ }\textbf
  {\bibinfo {volume} {38}},\ \bibinfo {pages} {187} (\bibinfo {year}
  {2004})}\BibitemShut {NoStop}%
\bibitem [{\citenamefont {Tanizawa}\ \emph {et~al.}(2005)\citenamefont
  {Tanizawa}, \citenamefont {Paul}, \citenamefont {Cohen}, \citenamefont
  {Havlin},\ and\ \citenamefont {Stanley}}]{Tanizawa:2005fd}%
  \BibitemOpen
  \bibfield  {author} {\bibinfo {author} {\bibfnamefont {T.}~\bibnamefont
  {Tanizawa}}, \bibinfo {author} {\bibfnamefont {G.}~\bibnamefont {Paul}},
  \bibinfo {author} {\bibfnamefont {R.}~\bibnamefont {Cohen}}, \bibinfo
  {author} {\bibfnamefont {S.}~\bibnamefont {Havlin}}, \ and\ \bibinfo {author}
  {\bibfnamefont {H.~E.}\ \bibnamefont {Stanley}},\ }\href@noop {} {\bibfield
  {journal} {\bibinfo  {journal} {Phys. Rev. E},\ }\textbf {\bibinfo {volume}
  {71}},\ \bibinfo {pages} {047101} (\bibinfo {year} {2005})}\BibitemShut
  {NoStop}%
\bibitem [{\citenamefont {Tanizawa}\ \emph {et~al.}(2006)\citenamefont
  {Tanizawa}, \citenamefont {Paul}, \citenamefont {Havlin},\ and\ \citenamefont
  {Stanley}}]{Tanizawa:2006dn}%
  \BibitemOpen
  \bibfield  {author} {\bibinfo {author} {\bibfnamefont {T.}~\bibnamefont
  {Tanizawa}}, \bibinfo {author} {\bibfnamefont {G.}~\bibnamefont {Paul}},
  \bibinfo {author} {\bibfnamefont {S.}~\bibnamefont {Havlin}}, \ and\ \bibinfo
  {author} {\bibfnamefont {H.~E.}\ \bibnamefont {Stanley}},\ }\href@noop {}
  {\bibfield  {journal} {\bibinfo  {journal} {Phys. Rev. E},\ }\textbf
  {\bibinfo {volume} {74}},\ \bibinfo {pages} {016125} (\bibinfo {year}
  {2006})}\BibitemShut {NoStop}%
\bibitem [{\citenamefont {Paul}\ \emph {et~al.}(2006)\citenamefont {Paul},
  \citenamefont {Sreenivasan}, \citenamefont {Havlin},\ and\ \citenamefont
  {Stanley}}]{Paul:2006cd}%
  \BibitemOpen
  \bibfield  {author} {\bibinfo {author} {\bibfnamefont {G.}~\bibnamefont
  {Paul}}, \bibinfo {author} {\bibfnamefont {S.}~\bibnamefont {Sreenivasan}},
  \bibinfo {author} {\bibfnamefont {S.}~\bibnamefont {Havlin}}, \ and\ \bibinfo
  {author} {\bibfnamefont {H.~E.}\ \bibnamefont {Stanley}},\ }\href@noop {}
  {\bibfield  {journal} {\bibinfo  {journal} {Physica A},\ }\textbf {\bibinfo
  {volume} {370}},\ \bibinfo {pages} {854} (\bibinfo {year}
  {2006})}\BibitemShut {NoStop}%
\bibitem [{\citenamefont {Newman}(2002)}]{Newman:2002jj}%
  \BibitemOpen
  \bibfield  {author} {\bibinfo {author} {\bibfnamefont {M.~E.~J.}\
  \bibnamefont {Newman}},\ }\href@noop {} {\bibfield  {journal} {\bibinfo
  {journal} {Phys. Rev. Lett.},\ }\textbf {\bibinfo {volume} {89}},\ \bibinfo
  {pages} {208701} (\bibinfo {year} {2002})}\BibitemShut {NoStop}%
\bibitem [{\citenamefont {Goltsev}\ \emph {et~al.}(2008)\citenamefont
  {Goltsev}, \citenamefont {Dorogovtsev},\ and\ \citenamefont
  {Mendes}}]{Goltsev:2008bf}%
  \BibitemOpen
  \bibfield  {author} {\bibinfo {author} {\bibfnamefont {A.~V.}\ \bibnamefont
  {Goltsev}}, \bibinfo {author} {\bibfnamefont {S.~N.}\ \bibnamefont
  {Dorogovtsev}}, \ and\ \bibinfo {author} {\bibfnamefont {J.~F.~F.}\
  \bibnamefont {Mendes}},\ }\href@noop {} {\bibfield  {journal} {\bibinfo
  {journal} {Phys. Rev. E},\ }\textbf {\bibinfo {volume} {78}},\ \bibinfo
  {pages} {051105} (\bibinfo {year} {2008})}\BibitemShut {NoStop}%
\bibitem [{\citenamefont {Schneider}\ \emph {et~al.}(2011)\citenamefont
  {Schneider}, \citenamefont {Moreira}, \citenamefont {Andrade~Jr},
  \citenamefont {Havlin},\ and\ \citenamefont {Herrmann}}]{Schneider:2011ip}%
  \BibitemOpen
  \bibfield  {author} {\bibinfo {author} {\bibfnamefont {C.~M.}\ \bibnamefont
  {Schneider}}, \bibinfo {author} {\bibfnamefont {A.~A.}\ \bibnamefont
  {Moreira}}, \bibinfo {author} {\bibfnamefont {J.~S.}\ \bibnamefont
  {Andrade~Jr}}, \bibinfo {author} {\bibfnamefont {S.}~\bibnamefont {Havlin}},
  \ and\ \bibinfo {author} {\bibfnamefont {H.~J.}\ \bibnamefont {Herrmann}},\
  }\href@noop {} {\bibfield  {journal} {\bibinfo  {journal} {Proc. Natl. Acad. Sci. USA},\ }\textbf
  {\bibinfo {volume} {108}},\ \bibinfo {pages} {3838} (\bibinfo {year}
  {2011})}\BibitemShut {NoStop}%
\bibitem [{\citenamefont {Herrmann}\ \emph {et~al.}(2011)\citenamefont
  {Herrmann}, \citenamefont {Schneider}, \citenamefont {Moreira}, \citenamefont
  {Andrade~Jr},\ and\ \citenamefont {Havlin}}]{Herrmann:2011hd}%
  \BibitemOpen
  \bibfield  {author} {\bibinfo {author} {\bibfnamefont {H.~J.}\ \bibnamefont
  {Herrmann}}, \bibinfo {author} {\bibfnamefont {C.~M.}\ \bibnamefont
  {Schneider}}, \bibinfo {author} {\bibfnamefont {A.~A.}\ \bibnamefont
  {Moreira}}, \bibinfo {author} {\bibfnamefont {J.~S.}\ \bibnamefont
  {Andrade~Jr}}, \ and\ \bibinfo {author} {\bibfnamefont {S.}~\bibnamefont
  {Havlin}},\ }\href@noop {} {\bibfield  {journal} {\bibinfo  {journal}
  {J. Stat. Mech.},\ }\textbf
  {\bibinfo {volume} {2011}},\ \bibinfo {pages} {P01027} (\bibinfo {year}
  {2011})}\BibitemShut {NoStop}%
\bibitem [{\citenamefont {Tanizawa}\ \emph {et~al.}(2012)\citenamefont
  {Tanizawa}, \citenamefont {Havlin},\ and\ \citenamefont
  {Stanley}}]{Tanizawa:2012hh}%
  \BibitemOpen
  \bibfield  {author} {\bibinfo {author} {\bibfnamefont {T.}~\bibnamefont
  {Tanizawa}}, \bibinfo {author} {\bibfnamefont {S.}~\bibnamefont {Havlin}}, \
  and\ \bibinfo {author} {\bibfnamefont {H.~E.}\ \bibnamefont {Stanley}},\
  }\href@noop {} {\bibfield  {journal} {\bibinfo  {journal} {Phys. Rev.
  E},\ }\textbf {\bibinfo {volume} {85}},\ \bibinfo {pages} {046109} (\bibinfo
  {year} {2012})}\BibitemShut {NoStop}%
\bibitem [{\citenamefont {Shiraki}\ and\ \citenamefont
  {Kabashima}(2010)}]{Shiraki:2010is}%
  \BibitemOpen
  \bibfield  {author} {\bibinfo {author} {\bibfnamefont {Y.}~\bibnamefont
  {Shiraki}}\ and\ \bibinfo {author} {\bibfnamefont {Y.}~\bibnamefont
  {Kabashima}},\ }\href@noop {} {\bibfield  {journal} {\bibinfo  {journal}
  {Phys. Rev. E},\ }\textbf {\bibinfo {volume} {82}},\ \bibinfo {pages}
  {036101} (\bibinfo {year} {2010})}\BibitemShut {NoStop}%
\bibitem [{\citenamefont {Tanizawa}(2013)}]{Tanizawa:2013fy}%
  \BibitemOpen
  \bibfield  {author} {\bibinfo {author} {\bibfnamefont {T.}~\bibnamefont
  {Tanizawa}},\ }\href@noop {} {\bibfield  {journal} {\bibinfo  {journal}
  {NOLTA, IEICE},\ }\textbf {\bibinfo {volume} {4}},\ \bibinfo {pages} {138}
  (\bibinfo {year} {2013})}\BibitemShut {NoStop}%
  \bibitem [{\citenamefont {menche}(2010)}]{Menche:2010as}
  \BibitemOpen
  \bibfield {author} {\bibinfo {author} {\bibfnamefont {J.}\ \bibnamefont
  {Menche}}, \bibinfo {author} {\bibfnamefont {A.}\ \bibnamefont {Valleriani}}, and\
  \bibinfo {author} {\bibfnamefont {R.}\ \bibnamefont {Lipowsky}},\ } \href@noop {}
  {\bibfield  {journal} {\bibinfo  {journal} {Phys. Rev. E},\ }
  \textbf {\bibinfo {volume} {81}},\ \bibinfo {pages} {046103}
  (\bibinfo {year} {2010})}\BibitemShut {NoStop}%
\bibitem [{\citenamefont {Litvak}(2013)}]{Litvak:2013un}
  \BibitemOpen
  \bibfield {author} {\bibinfo {author} {\bibfnamefont {N.}\ \bibnamefont
  {Litvak}}, and\ \bibinfo {author} {\bibfnamefont {R.}\ \bibnamefont {van~der~Hofstad}},\
  } \href@noop {} {\bibfield  {journal} {\bibinfo {journal} {Phys. Rev. E},\ }
  \textbf {\bibinfo {volume} {87}},\ \bibinfo {pages} {022801}
  (\bibinfo {year} {2013})}\BibitemShut {NoStop}%
\bibitem [{\citenamefont {Yang}(2016)}]{Yang:2016lo}
  \BibitemOpen
  \bibfield {author} {\bibinfo {author} {\bibfnamefont {D.}\ \bibnamefont
  {Yang}}, \bibinfo {author} {\bibfnamefont {L.}\ \bibnamefont {Pan}},
  and\ \bibinfo {author} {\bibfnamefont {T.}\ \bibnamefont {Zhou}},\
  } \href@noop {} {\bibfield  {journal} {\bibinfo  {journal} {arXiv:1602.04350}\
  }}\BibitemShut {NoStop}%
\bibitem [{\citenamefont {Newman}(2003)}]{Newman:2003fh}%
  \BibitemOpen
  \bibfield  {author} {\bibinfo {author} {\bibfnamefont {M.~E.~J.}\
  \bibnamefont {Newman}},\ }\href@noop {} {\bibfield  {journal} {\bibinfo
  {journal} {Phys. Rev. E},\ }\textbf {\bibinfo {volume} {67}},\ \bibinfo
  {pages} {026126} (\bibinfo {year} {2003})}\BibitemShut {NoStop}%
\bibitem [{\citenamefont {V{\'a}zquez}\ and\ \citenamefont
  {Moreno}(2003)}]{Vazquez:2003fq}%
  \BibitemOpen
  \bibfield  {author} {\bibinfo {author} {\bibfnamefont {A.}~\bibnamefont
  {V{\'a}zquez}}\ and\ \bibinfo {author} {\bibfnamefont {Y.}~\bibnamefont
  {Moreno}},\ }\href@noop {} {\bibfield  {journal} {\bibinfo  {journal}
  {Phys. Rev. E},\ }\textbf {\bibinfo {volume} {67}},\ \bibinfo {pages}
  {015101} (\bibinfo {year} {2003})}\BibitemShut {NoStop}%
  \bibitem [{\citenamefont {Colomer-de Sim\'{o}n}\ and\ \citenamefont
  {Bogu\~{n}\'{a}}(2014)}]{Colomer:2014}%
  \BibitemOpen
  \bibfield  {author} {\bibinfo {author} {\bibfnamefont {P.}~\bibnamefont
  {Colomer-de Sim\'{o}n}}\ and\ \bibinfo {author} {\bibfnamefont
  {M.}~\bibnamefont {Bogu\~{n}\'{a}}},\ }\href@noop {} {\bibfield  {journal}
  {\bibinfo  {journal} {Physical Review X},\ }\textbf {\bibinfo {volume} {4}},\
  \bibinfo {pages} {041020} (\bibinfo {year} {2014})}\BibitemShut {NoStop}%
  \bibitem [{\citenamefont {Noh}(2007)}]{Noh:2007cl}%
  \BibitemOpen
  \bibfield  {author} {\bibinfo {author} {\bibfnamefont {J.~D.}\ \bibnamefont
  {Noh}},\ }\href@noop {} {\bibfield  {journal} {\bibinfo  {journal} {Phys.
   Rev. E},\ }\textbf {\bibinfo {volume} {76}},\ \bibinfo {pages} {026116}
  (\bibinfo {year} {2007})}\BibitemShut {NoStop}%
\end{thebibliography}

%
\end{document}